\documentclass{article}%
\usepackage{amsmath}
\usepackage{amsfonts}
\usepackage{amssymb}
\usepackage{graphicx}
\usepackage{multirow}
\usepackage{fleqn}

\title{Numerical Experiments}
\author{Sverre Aarseth\\
Institute of Astronomy, University of Cambridge, U.K.}
\begin{document}

\maketitle{}

To the best of my knowledge, the phrase "Numerical Experiments" was coined by
Michel H\'enon himself in 1966, or as he called it "Experiences Num\'erique"
\cite{H67a}.
His idea was that one can study stellar systems within the context of
a laboratory where all the parameters are known.
Moreover, the effect of varying the parameters can be measured directly.
This point of view conveniently ignores the role of chaos in non-linear systems
but the principle still applies in a statistical sense. 
Bearing in mind the primitive state of the subject at the time, the new concept
was very perceptive and is now taken for granted.

Given my own long association with the $N$-body problem, it therefore makes a
fitting theme for these brief comments on related contributions which were mainly
made during his early career.
I first met Michel in 1964 during the IAU Symposium No. 25 in Thessaloniki which also
happened to be my first international meeting.
In those early days, the number of people interested in performing direct $N$-body
calculations was quite small so it was natural that we established close contact for
the future.
This contact was strengthened further by attending stellar dynamical conferences in
Besancon (1966) and Paris (1968).
Another notable event was that Michel was the first person to give a seminar at the
brand new Institute of Theoretical Astronomy in the spring of 1967 (renamed IoA from
1972).
On a personal level, I spent a few weeks at Nice Observatory around 1973
where I enjoyed fruitful discussions with Michel which resulted in a joint paper.
I also came to appreciate his understanding of numerical errors.
Thus in his opinion, the 1966 Bulirsch--Stoer method for numerical integration
achieved the maximum accuracy that can be attained on a given computer.
I also remember well his novel system for extracting specific references from a
card index.

The Ph.D. thesis of Michel H\'enon \cite{H61} has played a major role for understanding
the gravitational $N$-body problem as well as star cluster evolution.
A key point here is that clusters evolve in response to energy generation at the centre.
This novel idea anticipated the importance of binaries which was eventually clarified a
decade later.
In the proposed homology model, the evolution is towards infinite central density where
the energy is accumulated by a small number of stars.
To maintain energy conservation, about one-third of the energy is carried away by
escaping stars.
It is quite appropriate that an English translation of the two fundamental papers should
now be available 50 years later on the astro-ph website (\cite{H11a},\cite{H11b}).

In stellar dynamics, Michel was also a pioneer of the Monte Carlo method.
His first attempt may have been inspired by the so-called Shell method (\cite{H67a})
which provided a fast procedure for numerical study of spherical systems, where the
expensive force calculation was simplified.
These ideas eventually led to his Monte Carlo formulation (\cite{H67b}, \cite{H72a}, \cite{H72b}), especially
as presented at the Cambridge 1970 IAU Colloquium on the $N$-body problem 
(\cite{H72a}, \cite{H72b}) and perfected at the time of the IAU Symposium No. 69
\cite{H75}.
This work enabled the Monte Carlo method to be used beyond the singularity and further
refinement of the Coulomb factor which plays a fundamental role in relaxation
theory brought the results in closer agreement with the direct $N$-body method. 
In summary, the main idea of the H\'enon method is based on random selection of
pair-wise interactions while other groups have employed more traditional algorithms
where the particle orbits are integrated separately at additional cost.
During the intervening years, the Monte Carlo method has received much attention
such that current modelling essentially includes all the relevant astrophysical
processes and it also gives excellent agreement for core collapse.

The question of escape in self-gravitating systems has been addressed by many authors.
Inevitably, simplifying assumptions are introduced and this leads to very large
differences in escape rates and dissolution times.
This challenging topic was tackled by H\'enon in his 1969 paper \cite{H69} which provided
explicit solutions for an arbitrary initial mass function.
Two simplifying assumptions were employed: isotropic velocities and spherical symmetry.
For the first time, the contribution to the escape rate from different masses was
evaluated.
In this way, the importance of the mass spectrum was emphasized.
For example, 50 \% of the effect could be ascribed to a small number of the most
massive stars.
Based on these results, it became clear that realistic star cluster simulations
needed to include a wide range of masses with only a small fraction of heavy stars
playing a vital role.
Subsequent theoretical work by others also showed that the relaxation time as well as
the core collapse times were reduced significantly.
Although the available cluster membership was quite limited in the 1970s, the qualitative
agreement with simulations was found to be satisfactory and in particular pointed to
very small escape rates in the equal-mass case.
In his second classical 1961 paper, H\'enon analysed the process of escape from isolated
systems.
He showed convincingly that escape requires contributions from close encounters because
diffusion by itself would take an infinite time to be effective.

The reliability of numerical solutions based on direct methods was under discussion
from an early stage.
Such doubts were dispelled to some extent by the publication of the so-called comparison
paper \cite{AHW74} which compared evolution rates with those
obtained by faster methods; i.e. Monte Carlo and fluid dynamics.
More recently, impressive agreement was found for the core collapse times of larger
systems based on Monte Carlo and Fokker--Planck methods which therefore gave support
for H\'enon's confidence in the numerical approach.
The comparison paper also provided a detailed algorithm for constructing initial
conditions for the Plummer model which contains unmistakable traces of his mathematical
genius.
It goes without saying that many $N$-body simulations have benefited from this precise
recipe.

Michel H\'enon was also involved in starting pioneering work in dynamical chaos theory.
As has been discussed extensively in another contribution at this meeting,
the original paper \cite{H64} has had a profound influence and reflects a
deep insight obtained by studying relatively simple equations.
The restricted three-body problem is another subject that caught his attention
already in the 1960s in which he pursued another career.
Again he produced a series of single-authored papers on this classical problem which
is best left to others to evaluate.

The comments above are restricted to the subject of stellar dynamics where Michel made
lasting contributions.
Unlike some of his colleagues, he was versatile and continued to employ his considerable
mathematical talents in the quest for new challenges.
It can be said that he demonstrated the rare ability of combining theory and numerical
experiments in a masterful way.

\end{document}